# Why not only electrostatic discharge but even a minimum charge on the surface of highly sensitive explosives can catalyze their gradual exothermic decomposition and how a cloud of unipolar charged explosive particles turns into ball lightning

O. Meshcheryakov [*]

[*]Wing Ltd Company
33 French boulevard, Odessa 65000, Ukraine, wing99@mail.ru

## ABSTRACT

Even a single excess electron or ion migrating on the surface of sensitive explosives can catalyze their gradual exothermic decomposition. Mechanisms underlying such a charge-induced gradual thermal decomposition of highly sensitive explosives can be different. If sensitive explosive is a polar liquid, intense charge-dipole attraction between excess surface charges and surrounding explosive molecules can result in repetitive attempts of solvation of these charges by polar explosive molecules. Every attempt of such uncompleted nonequilibrium solvation causes local exothermic decomposition of thermolabile polar molecules accompanied by further thermal jumping unsolvated excess charges to new surface sites. Thus, ionized mobile hot spots emerge on charged explosive surface. Stochastic migration of ionized hot spots on explosive surface causes gradual exothermic decomposition of the whole mass of the polar explosive. The similar gradual charge-catalyzed exothermic decomposition of both polar and nonpolar highly sensitive explosives can be also caused by intense charge-dipole attacks of surrounding water vapor molecules electrostatically attracted from ambient humid air and strongly accelerated towards charged sites on explosive surfaces. Emission of thermoelectrons, photons and heat from ionized hot spots randomly migrating on charged surface of highly sensitive explosive aerosol nanoparticles converts such particles into the form of short-circuited thermionic nanobatteries.

## 1 SHORT-CIRCUITED NANOBATTERIES

The nature of ball lightning still remains mysterious, and numerous attempts to experimentally simulate this phenomenon remain inconclusive. We have suggested in [1,2] that certain classes of highly exothermic reactions, e.g., such as the charge-catalyzed water vapor induced redox reactions or also the reactions of the self-propagating high-temperature synthesis, occurring on the surface or in the volume of some kinds of unipolar charged combustible aerosol particles, can underlie the ball lightning phenomenon. Under certain conditions, such exothermic reactions can cause the self-oscillating high-frequency processes of periodical separation and relaxation of electric charges occurring either on surface or inside the burning aerosol nanoparticles. Correspondingly, such aerosol particles can be spontaneously getting the properties of periodically short-circuited tiny aerosol batteries, in which the primary energy of the intra-particle exothermic reactions directly converts into electromagnetic energy.

According to [1,2]: a) aerosol particles possesing the properties of the short-circuited aerosol batteries can be spontaneously synthesized in both the low-temperature and high-temperature processes, particularly, in high-voltage electric discharges; b) in the form of luminous smoke clouds such aerosol particles-batteries can be generated in high-voltage arc discharges due either to electrostatic/ plasma spraying of combustible electrode components or to the ion co-condensation of the arc-evaporated mutually reactive electrode substances; c) a ball- or thread- shaped luminous cloud - ball lightning - can be self-assembled of trillions of the unipolar charged particles-batteries owing to intense electromagnetic dipole-dipole attraction arising between the periodically short-circuited aerosol batteries, each of which at any point of time is an electric dipole, when short-circuiting being also a magnetic dipole; d) very different kinds of combustible nanoparticles can be spontaneously converted into the form of the short-circuited nanobatteries, involving different mechanisms of periodical separation and relaxation of electric charges inside these particles or on their surface; e) according to various possible mechanisms of the periodical separation and relaxation of electric charges inside the aerosol particles or on their surface, the different kinds of such particles can be spontaneously converted into the form of the electrochemical, thermoelectric, thermionic, pyroelectric, photoelectron emission, photoelectric, or radioisotope electric aerosol nanogenerators; f) in humid air, even minimally charged nanoparticles of many active metals can be spontaneously converted into the form of short-circuited metal-air nanobatteries due to the charge-catalyzed, predominantly water vapor induced electrochemical, i.e., ion-mediated oxidation occurring on the surface of the nanoparticles; g) many composite aerosol nanoaggregates, e.g., those containing soot carbon nanoparticles plus certain metallic, metal oxide or molten carbonate/hydroxide based nanocomponents, can be spontaneously converted into the carbon anode containing aerosol nanobatteries, in which highly-exothermic processes of electrochemical oxidation of carbon nanoparticles are accompanied by synchronous processes of intra-particle charge separation and relaxation;

h) thermoelectric nano/micro generators and also thermionic nano- or micro converters can be spontaneously formed as a result of highly exothermic reactions of self-propagating high-temperature synthesis, initiated within composite particles containing mutually reacting condensed nanocomponents; for example, such short-circuited thermoelectric and thermionic aerosol nanobatteries can be spontaneously formed of burning aerosol nanothermites or different mutually reacting nanocomposites; the rate of electrogenerating exothermic reactions within these composite nanobatteries can be practically independent of the rate of their ultra-decelerated oxidation by the external air oxidants, and thermoelectric/thermionic nanobatteries, based on SHS reacting aerosol nanocomposites, can be exceptionally self-contained objects; i) a high dissipative and aggregative stability of the ball lightning clouds self-assembled of the unipolar charged, periodically short-circuited aerosol particles-batteries is based on the dynamic equilibrium that arise in these clouds due to competition between the forces of the interparticle attraction and repulsion, in particular, due to dynamic competition between the long-range forces of the interparticle electromagnetic dipole-dipole attraction and the total forces of electrostatic, thermophoretic, and diffusiophoretic mutual interparticle repulsion; it is worth emphasizing here that the presence of electric currents inside the aerosol particles or on their surface always contributes to their intense mutual dipole-dipole attraction and even to their mutual coagulation; if the intra-particle electric currents are direct currents, such dipole-dipole attraction occurring between the current-carrying aerosol particles is purely magnetic dipole-dipole attraction; if the intra-particle electric currents are alternating currents, such dipole-dipole attraction occurring between these periodically current-carrying aerosol particles should be more correctly called as the "electromagnetic" dipole-dipole attraction, containing both components: electrostatic dipole-dipole attraction operating between non-short-circuited aerosol particles-batteries and magnetic dipole-dipole attraction that operate between these particles-batteries at the repetitive moments of their short-circuiting.

Thus, the repetitive processes of separation and relaxation of electric charges occurring within or on the surface of aerosol particles, allow us to consider these particles as periodically short-circuited tiny electric generators. Generally speaking, such periodic current processes connected with intraparticle separation and relaxation of electric charges can be fed not only by an energy primarily accumulated in the aerosol particles.

In particular, the primary energy can be stored within the aerosol particles in the form of a chemical, nuclear, magnetic energy of nondissipative supercurrents, etc. This pre-accumulated energy can be emitted from the surface of the nano- or submicron aerosol particles, both gradually and extremely nonisotropically, e.g., in the form of individual hot spots stochastically arising on the surface of the energy-releasing particles. Substantial energy release within the hot spots can be followed by a local thermionic emission. The local processes of intense emission of thermoelectrons from hot spots stochastically arising and migrating on the surface of the energy-releasing particle also can have random, chaotic character. Electrons readily "evaporating" from the hot spots on the particle surface will "condense" onto colder surface sites, continuously generating thermionic currents and discharges on the surface of the aerosol particles, and turning a cloud of such nonisotropically gradually energy-releasing aerosol particles into a self-compressing cloud of the electromagnetically mutually attracting short-circuited thermionic batteries.

It is clear that this nonisotropic character of thermionic emission from the surface of small energy-releasing aerosol particles is a consequence of both the nonuniform surface heating and small sizes of these particles, in these cases a number of electrons emitted from one side of a small particle is not equivalent to the number of electrons emitted from the opposite side of this particle for any sufficiently small interval of time.

According to [1,2], the luminous clouds consisting of trillions of the aerosol nanobatteries can possess all the contradictory properties attributed to ball lightning, namely: an ability to make impressive shape metamorphoses, an ability to achieve the high-speed, at the same time, deformation-free flying, also abilities to elastic bouncing on the solid surface, to smoke-like penetrating through small holes, to generating both intense optical and radio-frequency radiation, as well as an ability to explode suddenly with synchronous generating the extremely strong electromagnetic pulses.

Approximately half of the eyewitnesses of those who have directly observed the final phases of ball lightning life asserted that of disappearance of ball lightning was highly explosive and that the ball lightning explosions were frequently accompanied by various electromagnetic phenomena. In some cases the ball lightning explosions described were extremely powerful, resulting in the significant mechanical and electrical damage in the surrounding objects .

In our previous papers we have only discussed excess charge-catalyzed reactions of exothermic chemical synthesis, potentially leading to transformation of a cloud of combustible aerosol particles into electromagnetically self-assembled cloud consisting of the periodically short-circuited particles spontaneously getting the properties batteries. In this paper we would like to show that not only the exothermic reactions of chemical synthesis, such as charge-catalyzed electrochemical oxidation of the combustible nanoparticles or self-propagating high-temperature synthesis within high energetic nanocomposites, can contribute to spontaneous formation of the short-circuited electrochemical and/or thermionic aerosol nanobatteries.

A special type of the excess charge-catalyzed surface reactions – the excess electron, proton or ion -catalyzed gradual exothermic decomposition that can take place on the surface of highly charged aerosol particles consisting of extremely sensitive explosives could also contribute to converting such explosive aerosol particles into the form of short-circuited thermionic aerosol batteries.

Thus, further we would like to show that luminous clouds electromagnetically self-assembled of such unipolar charged gradually decomposed highly sensitive high explosive aerosol particles possessing the properties of the short-circuited thermionic aerosol batteries can constitute another widespread class of ball lightning. Atmospheric electric discharges can produce extremely sensitive high explosives in the form of the self-compressed luminous clouds consisting of unipolar charged explosive aerosol droplets or solid particles possessing the properties of the short-circuited thermionic aerosol batteries and subjected to the excess charge-catalyzed gradual exothermic decomposition. Powerful high-voltage spark or arc discharges from time to time occurring in electrical equipment, e.g., those arising as a consequence of external lightning surge, as well as the processes of the intense electrolysis occasionally taking place inside damaged wetted wiring are capable off generating small luminous clouds containing unipolar charged aerosol particles exposed to the excess charge-catalyzed gradual surface decomposition and consisting e.g. of the following extremely sensitive explosives in situ discharge-synthesized: silver nitride; silver azide; silver fulminate; silver acetylide, mixed polyynides, including silver polyynides; copper acetylides, including copper polyynides; copper fulminate; tetramine copper salts, e.g., those produced electrolytically; copper azide, copper amine azide; iron azide; zinc azide; lead azide; lead styphnate; lead picrate, carbonyl diazide, and also azides, picrates, and styphnates of some other metals less often used in electrical equipment in the form of metal contacts, solders or coating.

Additionally, small aerosol clouds consisting of the unipolar charged submicron or nanoparticles-batteries consisting of discharge-synthesized highly sensitive organic peroxide / ozonides based explosives, exposed to the excess charge-catalyzed gradual exothermic decomposition can be readily generated in the electrical equipment due to arc-induced pyrolytic oxidative depolymerization of some commonly used insulating materials, such as polystyrene, teflon, butadiene-styrene rubber, etc, with the formation of highly sensitive high explosive polymeric styrene peroxides butadiene peroxides, tetrafluorethylene peroxides etc.

Clouds consisting of the unipolar charged extremely sensitive high explosive aerosol particles can be spontaneously synthesized in the free air during both thunderstorms and under fair weather conditions.

Such clouds can contain highly charged micrometer-sized, submicron or nano particles/ droplets consisting of in-situ synthesized condensed-phase highly sensitive high explosives. The excess ions, electrons or holes thermally migrating on the surface of such unipolar charged highly sensitive explosive particles can catalyze their gradual exothermic decomposition due to local charge-dipole destabilization of these thermolabile molecules. The atmospheric electric discharges are especially efficient natural generators of the unipolar charged highly sensitive explosive aerosols. There are several different processes involved in the discharge-induced or spontaneous generation of highly sensitive high explosive aerosol particles in the atmosphere. Such natural processes to produce high explosive aerosol clouds in open air can be based on: a) biogenic synthesis, for example, a large-scale biogenic oceanic production of alkyl nitrates, particularly, of methyl nitrate; b) high-voltage discharge -induced nitration, with the final generating the high explosive nitro compounds; c) lightning or solar radiation -induced photochemical oxidation, including atmospheric gas-phase synthesis of high explosive ozone, hydrogen peroxide, peroxyacetyl nitrate, organic ozonides, etc; d) ozonization, including atmospheric synthesis of highly sensitive high explosive chlorine oxides, dinitrogen pentoxide, organic hydroperoxides, peroxides, superoxides, ozonides, etc (it is worth noting that aerosol particles of, first of all, organic ozonides are the most important and common candidates to be a material aerosol basis for the high explosives based ball lightning described here - an important additional property of many organic ozonides, commonly synthesized in gas-phase reactions of olefines and aromatic compounds with ozone, is their ability to react violently with the surrounding molecules of water vapor, which is potentially contributing not only to their excess charge-catalyzed surface exothermic thermolysis, but also to their additional exothermic hydrolysis in humid air – so, the lower the temperature and humidity of ambient air, the longer the lifetime of organic ozonide based ball lightning clouds; highly sensitive high explosive oligomeric aerosol products of ozonolysis of the coal- or oil- derived aromatic hydrocarbons, such as benzene triozonide, toluene ozonide, xylene ozonide, and particularly, coal- derived naphthalene ozonide and phenanthrene ozonide are major candidates to be aerosol components of high explosive ball lightning; e) super-high-voltage pulse implantation of the hydrated positive gaseous ions of containing nitrogen, e.g., such as $N_2^+ \cdot (H_2O)$ or $NH_3^+ \cdot (H_2O)$ into some metal surfaces, accompanied by the surface formation and next electrostatic dispersing of the highly sensitive high explosive azides or nitrides, e.g. $AgN_3$. The efficient and fast nitration of organic aerosol contained compounds, can occur due either to the high-voltage arcs/ positive streamers -generated hydrated gaseous ions of nitronium, $NO_2^+ \cdot (H_2O)_n$, or with the participation of the high-voltage discharge-generated molecules of dinitrogen pentoxide, a low-boiling explosive product of the reaction between $NO_2$ or $N_2O_4$ with ozone. Both these extremely powerful nitrating agents, i.e., the hydrated nitronium air ions and dinitrogen pentoxide, could be responsible for high-voltage discharge-induced synthesis of aerosols containing sensitive high explosive nitro compounds. Many of the above atmospheric reactions, in which highly sensitive high explosive aerosols could be readily synthesized, are exothermic. Therefore, either gas-phase or heterogeneous synthesis of the above highly sensitive explosives in the form of tiny aerosol particles possessing extremely highly developed surface and subjected to spontaneous quenching in relatively cold ambient air, creates significant thermodynamic advantages, providing the efficient heat sink in the above high explosive generating reactions. So, luminous clouds electromagnetically self-assembled of unipolar charged highly sensitive explosive aerosol particles subjected to excess charge-catalyzed gradual exothermic surface decomposition and continuously emitting and re-adsobing thermoelectrons, photons and heat from their developed surface, can be frequently synthesized in the atmosphere, particularly in the thunderstorm atmosphere, thereby producing mysterious phenomena called "ball lightning".

The excess charge catalyzed gradual exothermic decomposition of the unipolar charged high explosive aerosol particles, continuously emitting thermoelectrons from ionized hot spots randomly migrating on their surface, converts these explosive aerosol particles into the form of the unipolar charged short-circuited thermionic aerosol batteries mutually attracting due to intense electromagnetic dipole-dipole attraction.

The local charge-catalyzed exothermic decomposition gradually migrating on the relatively cold surface of the explosive aerosol particles are an isothermal pre-explosive process due to metastable thermal equilibrium arising between the total intra-cloud heat release and the total heat emission out from this cloud. In these cases the ball lightning does not explode. However, external fluctuations of air humidity, temperature, magnetic or electric fields are able to substantially redistribute local concentrations of the heat-generating aerosol particles in the ball lightning cloud. In these cases, such ball lightning clouds can explode, turning into 3D detonating high explosive aerosol systems.

## 2 EXPLOSIVE HYDRATION AND SOLVATION OF EXCESS ELECTRONS OR IONS THERMALLY MIGRATING ON THE SURFACE OF HIGHLY SENSITIVE HIGH EXPLOSIVE AEROSOL PARTICLE

The charge-dipole attraction and strong acceleration of surrounding polar gas molecules of ambient water vapor towards excess charges on the surface of charged highly sensitive explosive aerosol particle cause high-energy epithermal collisions of these polar gas molecules electrostatically accelerated on the length of their free path before their collision with the charged explosive surface to collision energies of ~ 1.6 eV - in the case of hydration of excess surface electron,, or to the collision energies of ~ 7.18 eV - in the case of hydration of excess surface proton, or to the collision energies of ~ 0.5 - 2.0 eV - in the cases of hydration of the majority of different excess surface ions,with corresponding repetitive attempts of the explosive nonequilibrium hydratation of these excess surface charges by the neighbouring molecules of water vapor.

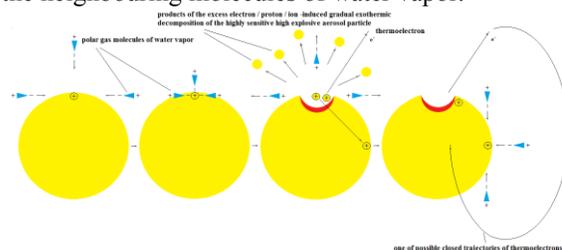

Fig. 1. Every highly exothermic event of the local explosive hydration/ solvation of excess surface electrons or ions by the surrounding water vapor molecules is accompanied by thermionic and photon emission from the ionized hot spot. The primary excess ion jumps to a new site on the highly sensitive explosive surface, where it attracts new surrounding molecules of water vapor or highly polar liquid explosive, and the process of explosive hydration / solvation of this excess charge is repeated again and again. Thus, the excess charges migrating on a surface of highly sensitive explosives, both the polar and nonpolar ones, can catalyze their gradual exothermic decomposition in humid air just due to the persistent high-energy charge-dipole collisions of the surrounding polar molecules of water vapor with the charged explosive surface.

Given the typical concentrations of water vapor molecules in normal air ( ~ 2 - 3 mole percent), it is easy to calculate that the factor of the charge-dipole epithermal acceleration of the water vapor molecules towards the minimally charged aerosol nanoparticles of highly sensitive explosives with typical activation energy of their exothermic decomposition ~ 45 - 160 kJ/mol, i.e., ~ 0.45 - 1.6 eV per one molecule, can cause the excess surface charge-induced gradual exothermic decomposition of such particles. In this case, all the explosive aerosol particles constituting "average" ball lightning with a diameter of about twenty centimeters, with a total net electric charge of about 1 microcoulomb, and with a minimum weight of the energy-generating condensed-phase explosive aerosol component of about four grams [1], can undergo the complete thermal decomposition for a period of ~ 100 seconds up to 1000 seconds, which is in good agreement with the typical lifetime of ball lightning.